\documentclass[runningheads]{llncs}
\usepackage{graphicx}
\usepackage{caption}
\usepackage{subcaption}
\usepackage{subscript}
\usepackage{mathtools}
\usepackage{gensymb}
\usepackage{placeins}
\usepackage{xcolor,soul}
\sethlcolor{lightgray}
\usepackage{pmat}
\usepackage{amssymb}

\begin{document}
\frontmatter          
\pagestyle{headings}  

\mainmatter              
\title{UV Exposed Optical Fibers with Frequency Domain Reflectometry for Device Tracking in Intra-Arterial Procedures}

\author{Francois Parent\inst{1} \and Maxime Gerard\inst{2} \and Raman Kashyap\inst{1} \and Samuel Kadoury\inst{2,3} }
 
\institute{Fabulas Lab, Dept. of Eng. Physics, Polytechnique Montreal, Montreal, Canada \and MedICAL Lab, Polytechnique Montreal, Montreal, QC, Canada  \and CHUM Research Center, Montreal, QC, Canada}
\authorrunning{Parent et al.}   

\titlerunning{Accepted to MICCAI 2017}

%

\maketitle              

\begin{abstract}

Shape tracking of medical devices using strain sensing properties in optical fibers has seen increased attention in recent years. In this paper, we propose a novel guidance system for intra-arterial procedures using a distributed strain sensing device based on optical frequency domain reflectometry (OFDR) to track the shape of a catheter. Tracking enhancement is provided by exposing a fiber triplet to a focused ultraviolet beam, producing high scattering properties. Contrary to typical quasi-distributed strain sensors, we propose a truly distributed strain sensing approach, which allows to reconstruct a fiber triplet in real-time. A 3D roadmap of the hepatic anatomy integrated with a 4D MR imaging sequence allows to navigate the catheter within the pre-interventional anatomy, and map the blood flow velocities in the arterial tree. We employed Riemannian anisotropic heat kernels to map the sensed data to the pre-interventional model. Experiments in synthetic phantoms and an $in$ $vivo$ model are presented. Results show that the tracking accuracy is suitable for interventional tracking applications, with a mean 3D shape reconstruction errors of $1.6 \pm 0.3$ mm. This study demonstrates the promising potential of MR-compatible UV-exposed OFDR optical fibers for non-ionizing device guidance in intra-arterial procedures.

\end{abstract}

\section{Introduction}

Intra-arterial therapies, such as trans-arterial chemoembolization (TACE), are now the preferred therapeutic approach for advanced hepatocellular carcinomas (HCCs). However, real-time localisation of the catheter inside the patient's vascular network is an important step during  embolizations, but remains challenging, especially in tortuous vessels and narrow bifurcations.vTraditional tracking approaches present a number of limitations for TACE, including line-of-sight requirements and tracking of flexible tools using infrared cameras, while workflow hinderances or metallic interferences are linked with electromagnetic (EM) tracking. Therefore alternative technologies have attempted to address these issues. A recent example uses bioimpedance models using integrated electrodes \cite{fuerst2016bioelectric} which infers the internal geometry of the vessel and mapped to a pre-interventional model, but is limited to the catheter tip. Optical shape sensing (OSS) is another technology measuring light deflections guided into optical fibers in order to measure strain changes in real-time, thereby inferring the 3D shape of the fiber by means of an integrative approach. Fiber Bragg grating (FBG) sensors can be integrated into submillimeter size tools, with no electromagnetic interference. Medical devices have incorporated FBGs in biopsy needles \cite{Park10}, catheters and other minimally invasive tools for shape detection and force sensing capabilities \cite{Elayaperumal14,Roesthuis14}. However FBGs only provide  discrete measurements, are costly to fabricate and reduce the flexibly of highly bendable tools. Optical frequency domain reflectometry (OFDR) is an alternative interferometric method with truly distributed sensing capabilities, frequently used to measure the attenuation along fibers. Duncan et al. compared the FBG and OFDR strain sensing approaches for optical fibers, showing an accuracy improvement with OFDR \cite{duncan2007high}. An array of 110  equally distanced FBGs was used, yielding an accuracy of 1.9mm, in comparison to a 3D shape reconstruction accuracy of 0.3mm using OFDR. Loranger et al. also showed that Rayleigh scattering, which is the basis of strain measurements using OFDR, can be considerably enhanced by exposing fibers to a UV beam, leading to an increase in backscattered signal by a factor of 6300  \cite{loranger2014rayleigh}.

\begin{figure}
  \centering
  \includegraphics[height=0.75in] {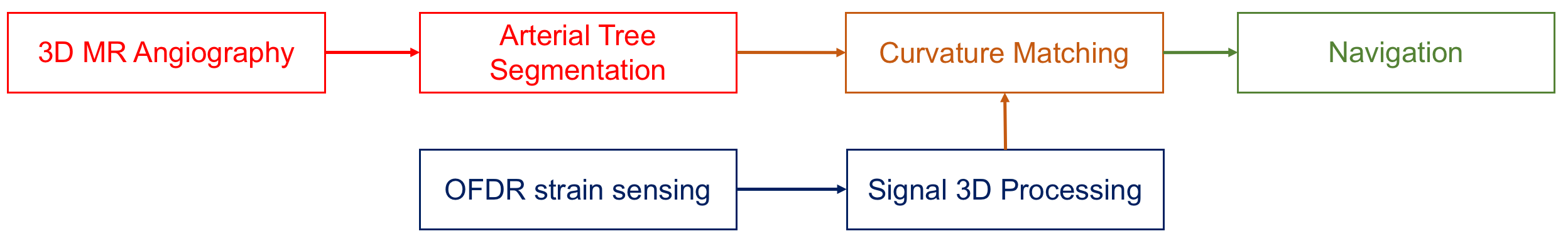}
  \caption{In OFDR navigation, strain measurements from UV exposed fibers with Rayleigh scattering are processed to 3D coordinates, which are mapped in real-time to an arterial tree model from a pre-interventional MR angiography.}
\label{fig:Workflow}
\end{figure}

In this paper, we present a new paradigm in catheter tracking using high scattering of a UV exposed fiber triplet inserted within a double-lumen catheter to perform real-time navigation in the hepatic arteries (Fig. \ref{fig:Workflow}). A custom made benchwork was first used to assemble three fibers in an equidistant geometry. In the proposed system, OFDR is based on Rayleigh scattering, which is caused by a random distribution of the refractive index on a microscopic scale in the fiber core of UV-doped optical fibers. The 3D shape of the fiber triplet was reconstructed according to the strain values measured by OFDR, and it's accuracy was evaluated both $in$ $vitro$ and $in$ $vivo$ to determine the catheter's tracking capabilites. In order to navigate the catheter within a patient's arterial tree, a 3D roadmap is automatically extracted from a 4D-flow MR imaging sequence, providing both anatomical and physiological information used for guidance in super-selective TACE procedures. Mapping between the sensed catheter shape and the anatomy is achieved using anisotropic heat kernels for intrinsic matching of curvature features. Rayleigh scattering processing has been proposed to obtain temperature measurements \cite{song2014long} and estimate strain properties in \cite{loranger2014rayleigh} but, to our knowledge, has not been applied to interventional navigation. The relative ordering of curvatures features (e.g. bifurcations) of the pre-operative models with the sensed strain values is not affected using dense intrinsic correspondences.

\section{Materials and Methods}
\label{sec:methods}

\subsection{Fabrication of UV enhanced optical fibers}

The proposed catheter is composed of three hydrogen loaded SMF-28 optical fibers (each with a 125$\mu$m diameter), exposed  to a focused UV beam (UVE-SMF-28). In our system, three fibers are glued together in a triangular geometry set apart by 120$^{\circ}$ (Fig. \ref{fig:Geometry}), using UV curing glue. Once the fibers are glued together, the outer diameter is approximately 260$\mu$m. The reusable and sterilizable fiber triplet was incorporated into a 0.67-mm-inner-diameter catheter (5-French Polyamide catheter, Cook, Bloomington, IN). 

\subsection{3D shape tracking using OFDR}

\begin{figure}
  \centering
  \includegraphics[height=1.8in] {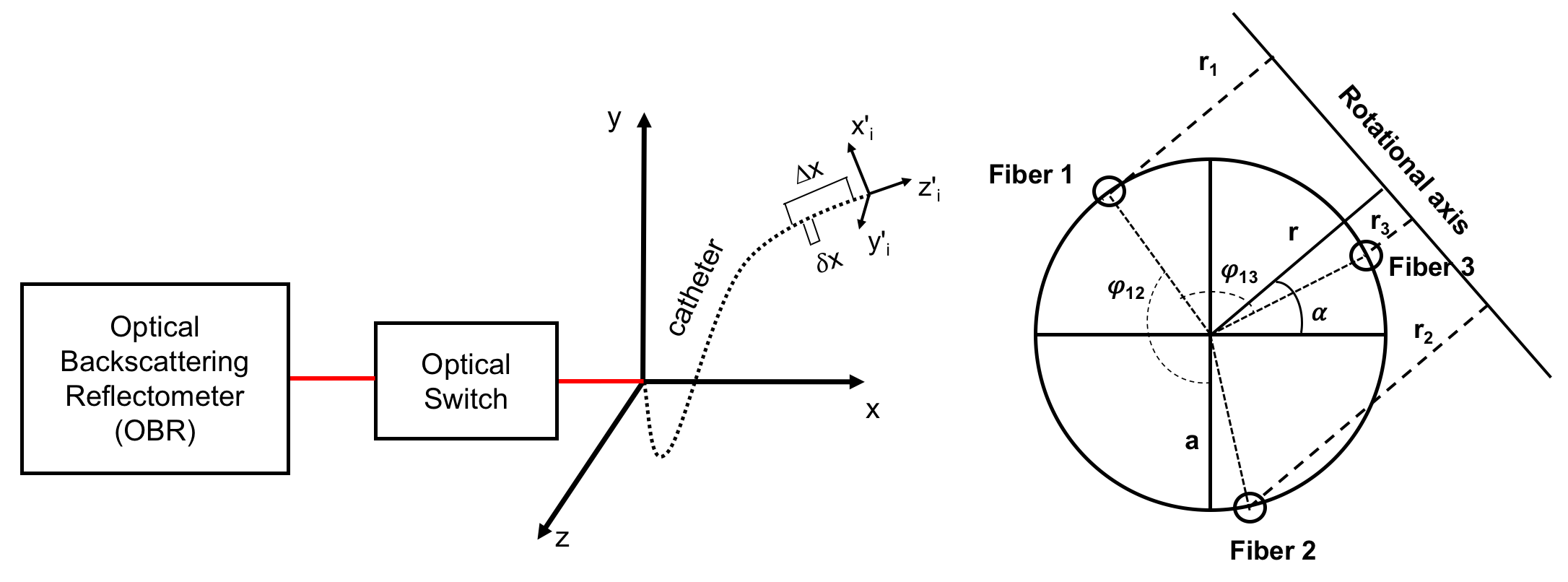}
  \caption{Diagram of the  optical systems used during measurements of the fiber triplet catheter. Illustration of a catheter separated in $i$-segments. Each segment defined within its own ($x_i^{'}$,$y_i^{'}$,$z_i^{'}$) frame can then be expressed in the tracking space $(x,y,z)$. The cross-section of the triplet of radius $a$ shows the angle between $x_i^{'}$ and the rotational axis, the distance between the center of the fiber triplet $r_i$,  the angle offset $\alpha_i$, as well as the angle $\varphi$ between each fiber.}
\label{fig:Geometry}
\end{figure}

The shape of the catheter is tracked using an OFDR method, which uses a frequency swept laser to interrogate the three fibers under test (FUT), successively. The backscatter signal of each FUT is then detected and analyzed in the frequency domain.  By using interferometric measurements, the strain along the fibers can be retrieved. A Fast Fourier Transform (FFT) is performed to evaluate the intensity of the backscatter signal as a function of the position along the fiber under test. Small-scale sections (corresponding to the spatial resolution $\Delta x$ of the strain sensor) of this signal is selected by an inverse FFT to evaluate the frequency response of this specific section. By comparing this frequency response of the fiber under strain and with the unstrained fiber, the local strain can be determined. To do so, a cross-correlation of the strained and unstrained spectra is performed. The corresponding cross-correlation spectra allows to precisely evaluate the spectrum drift between the reference and the measured section in the selected fiber length. The spectral drift is proportional to the strain (or temperature), so that the local strain or temperature can be calculated easily. In order to obtain a truly-distributed strain sensor, this process is repeated for each section of the FUT, successively. After selecting the desired length and location of the FUT, the desired fiber length (spatial resolution) ($\Delta x$) and the sensor spacing ($\delta x$), an optical backscattering reflectometer (OBR) provides distributed strain values along the desired region of the FUT as shown in Fig. \ref{fig:Geometry}.

Once OFDR is performed to evaluate the strain distributed along each fiber, a geometrical model proposed by Froggatt et al. \cite{froggatt1998high} is used to evaluate the position of the fiber triplet in tracking space. The core idea is to divide the triplet into segments $i$ and evaluate the position of the segments in its own frame ($x_i^{'}$,$y_i^{'}$,$z_i^{'}$). We use geometrical assumption to find the angle ($\alpha_i$) between the $x_i^{'}$ axis and the rotational axis of this segment, as  shown in Fig.  \ref{fig:Geometry}. Assuming $a_{ij}$ is the distance between the triplet center and the core of fiber $j$, $\varphi_{ijk}$ is the angle between each fiber core  $j$ and $k$ ($j$ and $k = \{1,2,3\}$, $k \neq j$) and $r_i$ is the distance between the triplet center and the rotational axis of this segment, the angle offset $\alpha_i$ and radius $r_i$ of the triplet can be obtained. The curvature and position of the segment tip in its own frame ($x_i^{'}$,$y_i^{'}$,$z_i^{'}$) can be evaluated. By applying a succession of projections and using rotational matrices, one can express these results in the laboratory frame ($x_i$,$y_i$,$z_i$) to reconstruct the entire 3D shape at a time $t$.  For more details see \cite{froggatt2010fiber}.

\subsection{Roadmapping of hepatic arteries }
Prior to navigation, the hepatic arterial tree used to map the sensed catheter shape and location onto the patient's anatomy is obtained through a segmentation algorithm that allows for the extraction of a complete 3D mesh model from a contrast-enhanced MR angiography (MRA) \cite{BADOUAL}.  The algorithm automatically detects the aorta and celiac trunk using an elliptical Hough transform following vesselness filtering on the MRA. An initial cylindrical triangular mesh is created around the detected aorta and deformed to fit the walls of the arteries by minimizing the  energy equation $E_{total} = E_{ext} + \beta E_{int}$. The first term represents the external energy driving the deformation of the mesh towards the edges of the vessel, using the magnitude of the intensity gradient vectors on the image. This term drives the triangles' barycenters towards their most promising positions. The second term in is the internal energy $E_{int}$. It limits the deformation by introducing topological constraints to ensure surface coherence, by measuring the neighbourhood consistency between the initial and optimized meshes. Finally, $\beta$ is a constant which allows for control of the trade-off between flexibility for deformation and surface coherence.  Each step of the iterative propagation consists in (a) duplicating a portion of the mesh extremity and translating it to that extremity, (b) orienting it by maximizing the gradient intensity values at its triangles barycenters, and (c) deforming it using the energy term.  A multi-hypothesis vessel tracking algorithm is used to detect bifurcations points and vessel paths to guide the adaptation process, generating the paths in the arterial tree and yielding a complete arterial model denoted as $C_{MRA}$. In addition to the earlier arterial phase contrast imaging, a 4D Flow imaging sequence was performed using a flow-encoded gradient-echo sequence with retrospective cardiac triggering and respiratory navigator gating. 

\subsection{Anisotropic curvature model matching}

Given a sensed catheter shape at time $t$ and the pre-operative roadmap $C_{MRA}$, the tracked catheter is then mapped to the patient-specific arterial model. We take advantage of the highly accurate curvature properties of the vascular tree to achieve shape correspondance, by using anisotropic heat kernels which are used as weighted mapping functions, enabling to obtain a local description of the intrinsic curvature properties within a manifold sub-space \cite{boscaini2016anisotropic}. We use an intrinsic formulation where the points are expressed only in terms of the Riemannian metric, which are invariant to isometric (metric-preserving) deformations. Prior to navigation, the 3D hepatic artery mesh model is divided in triangulated regions, which are defined by their unit normal vectors and principal curvature directions. Discretized anisotropic Laplacian sparse matrices are defined for each of these triangles, which include mass and stiffness matrices describing the anisotropic scaling and the rotation of the basis vector around the normal.  Once the arteries are expressed in spectral curvature signatures, it can be directly matched in real-time with the sensed OFDR data,  compensating for respiratory motion. 

\section{Experiments and Results}
\label{sec:results}

\subsection{Experimental setup}

The data processing is performed by an Optical Backscattering Reflectometer (OBR4600, LUNA Inc.). The sampling rate was determined based on the system's maximal capacity (1Hz), and an optical switch (JDSU SB series; Fiberoptic switch) with a channel transition period of 300ms was used to scan each fiber of the triplet, which were each exposed with a focused UV beam (Fig. \ref{fig.materials}a) during fabrication. Further data processing for catheter shape reconstruction considering the triplet characteristics was done by our own navigation software. 

\subsection{Synthetic vascular models}

\begin{figure}[t!]
\centering
  \includegraphics[width=4.7in] {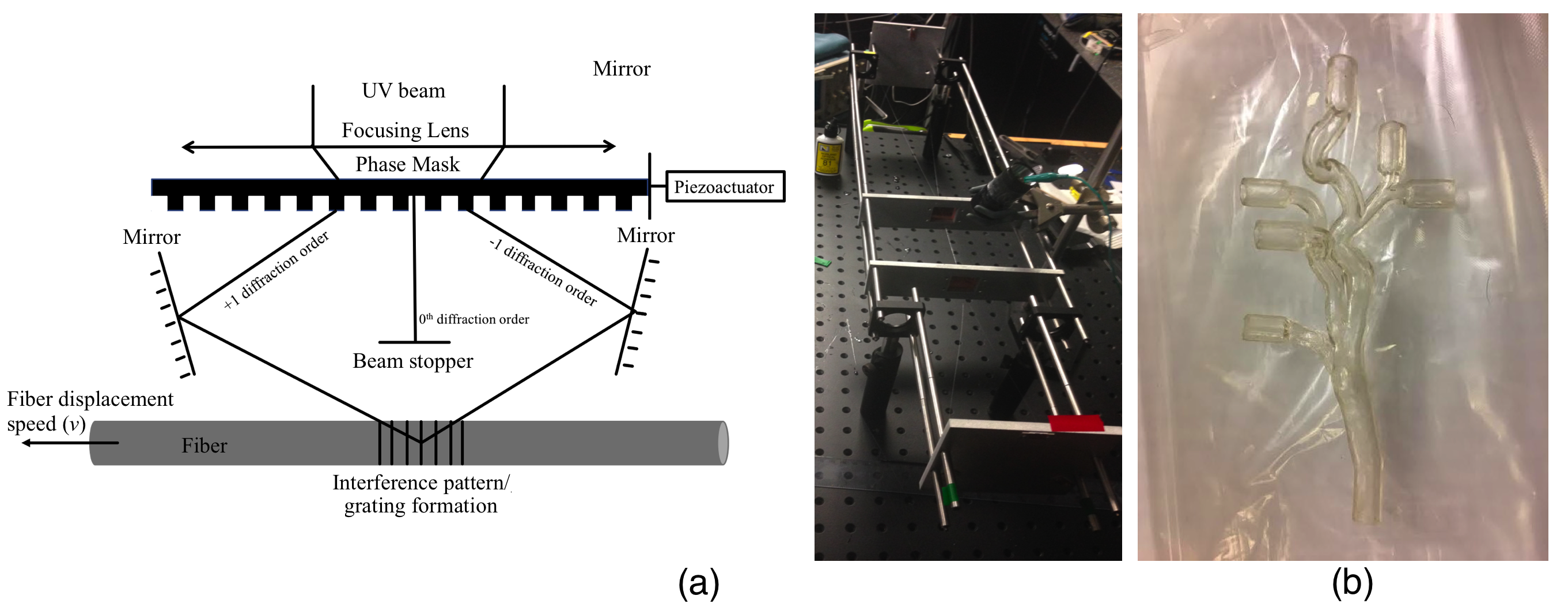}
\caption{(a) Fabrication setup with benchtest used only once (outside the clinic) for exposing focused UV beam for Rayleigh scattering on fiber triplet. (b) Example of a synthetic arterial phantom used for in vitro navigation for tracking accuracy assessment.}
\label{fig.materials}
\end{figure}

A set of 5 synthetic phantoms, created from stereolithography of patient-specific MRA's as shown in Fig. \ref{fig.materials}b were used to perform $in$ $vitro$ experiments inside an MR-scanner. The catheter was guided to a pre-defined target within the second segmental branch of the hepatic arterial tree on the MRA. Both tip position accuracy (Euclidean distance between virtual and physical tip) and root-mean-square differences (RMS) in the 3D catheter shape (15cm in length) were measured between a confirmation scan and the registered sensed data. Results were compared to EM tracked data, as shown in Table 1. The 3D shape RMS error was obtained by calculating the average point-to-point distances from a series of equidistant points taken along the virtual 3D shape to the closest point on the actual catheter. Compared to previous reports on FBG tracking \cite{mandal2016vessel}, these results show that the navigation accuracy is reliable, while remaining insensitive to MR magnetic fields. We also tested the tracking accuracy by measuring the amplitude of the backscatter signal from three types of fibers, which are standard single mode fiber (SMF-28), Germanium-boron doped fiber (Redfern) and hydrogen-loaded SMF-28 exposed to a focused UV beam (UVE-SMF-28). The UVE-SMF-28, which has a backscatter signal 6300 times higher than SMF-28, sees an average enhancement of 39\%, reaching 47\% for a highly curved regions in the phantom. The best accuracy was reached with the UVE-SMF-28, with an average tip accuracy of $1.1 \pm 0.4$mm and 3D shape error of $1.6 \pm 0.3$mm.
\begin{table*}[!t]
\renewcommand{\arraystretch}{1.3}
\caption{Target registration errors (tip accuracy and 3D root-mean-square (RMS) difference in shape) of the UV-enhanced shape sensing (UVE-28) catheter inside the MR gantry. Results are compared to electromagnetic (EM) tracking (Aurora, NDI), as well as to unenhanced optical fibers (SMF-28, Redfern).}
\label{table_example}
\centering
\scalebox{1}{
\begin{tabular}{|l||c|c|c|c|c|c|c|c|}
\hline

model \# & \multicolumn{4}{c|}{Tip accuracy (mm)} & \multicolumn{4}{c|}{3D RMS shape (mm)}  \\
    
\cline{2-9}
 & EM & SMF-28 & Redfern & \textbf{UVE-28} & EM & SMF-28 & Redfern & \textbf{UVE-28}  \\
\hline
1 &  $5.2 \pm 1.6$ &  $2.6 \pm 0.7$ &  $1.8 \pm 0.5$ &     $1.1 \pm 0.3$ &  $7.2 \pm 2.1$ & $3.0 \pm 0.7$ & $2.2 \pm 0.5$ &  $1.5 \pm 0.3$    \\
2 &  $8.7 \pm 2.0 $ &  $3.0 \pm 0.8$ &  $2.2 \pm 0.6$ &     $1.5 \pm 0.4$ &  $9.9 \pm 2.3$ & $3.5 \pm 0.8$ &  $2.7 \pm 0.6$ &  $1.9 \pm 0.4$   \\
3 &  $7.1 \pm 1.3$  & $2.5 \pm 0.7$ & $ 1.7 \pm 0.5$ &     $0.9 \pm 0.3$ &  $9.2 \pm 1.8$ & $2.9 \pm 0.6$  &  $2.1 \pm 0.4$ &  $1.4 \pm 0.2$ \\
4 &  $6.0 \pm 1.3$ &  $2.2 \pm 0.6$ & $ 1.5 \pm 0.4$ &     $0.8 \pm 0.3$ &  $7.5 \pm 1.7$ & $3.1 \pm 0.9$   &  $2.3 \pm 0.6$ &  $1.5 \pm 0.3$ \\
5 &  $8.1 \pm 1.9$ & $2.8 \pm 1.0$  &  $1.8 \pm 0.6$ &      $1.0 \pm 0.5$ & $8.6 \pm 1.9$ & $3.3 \pm 0.9$    & $2.5 \pm 0.7$ &  $1.8 \pm 0.4$\\
\hline
\textbf{Overall} &  $7.0 \pm 1.6$  & $2.6 \pm 0.7$    &   $1.8 \pm 0.5$ &        $1.1 \pm 0.4$ &  $8.5 \pm 2.0$ & $3.2 \pm 0.8$     &  $2.4 \pm 0.5$  & $1.6 \pm 0.3$           \\
 \hline
\end{tabular}
}
\end{table*}

\subsection{Animal experiment}
\begin{figure}[t!]
  \begin{minipage}[b]{0.46\linewidth}
  \centering
  \includegraphics[height=1.6in, width=2.3in] {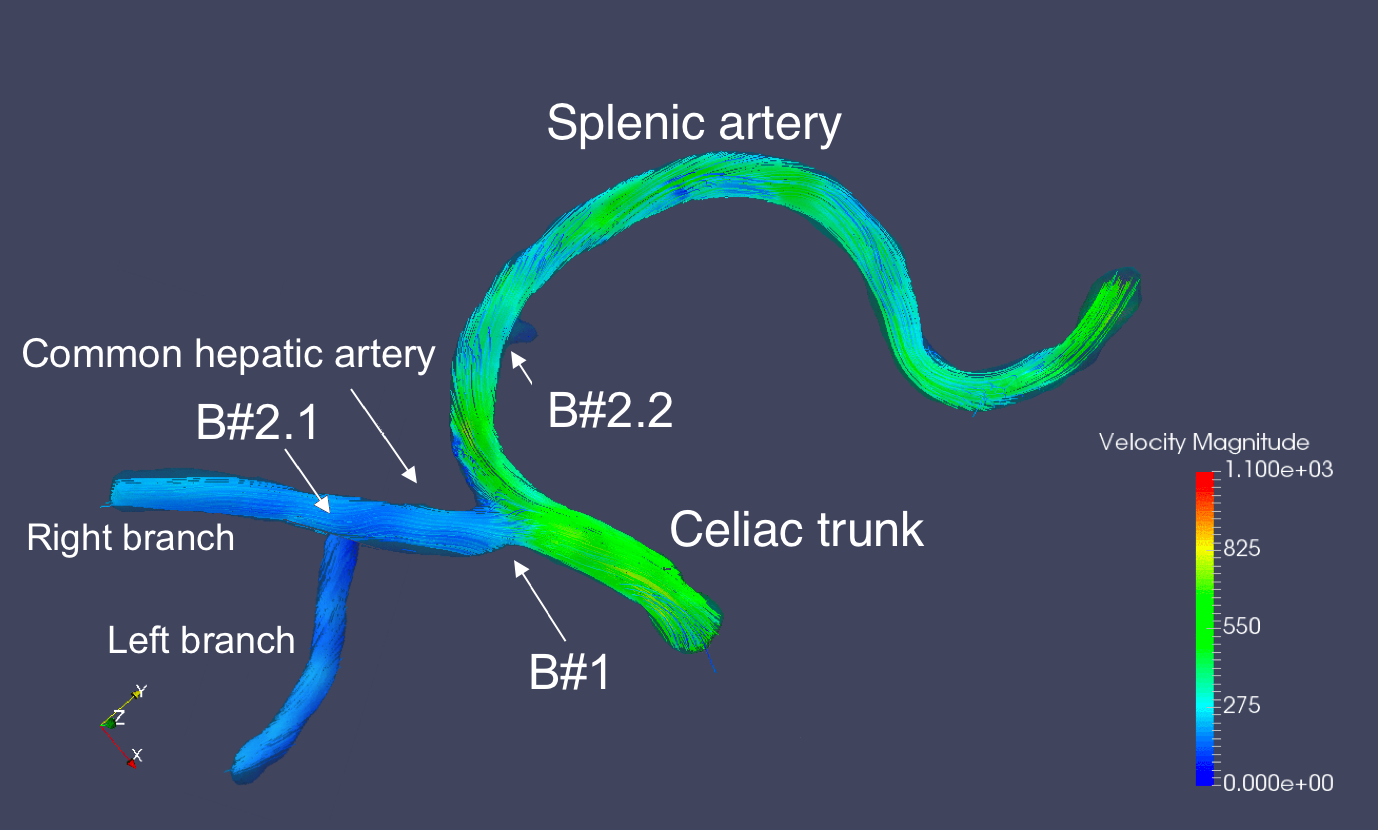}
  \vspace{0.2cm}
  \centerline{  (a)}
\end{minipage}
\begin{minipage}[b]{0.49\linewidth}
  \centering
  \includegraphics[height=1.7in, width=2.6in] {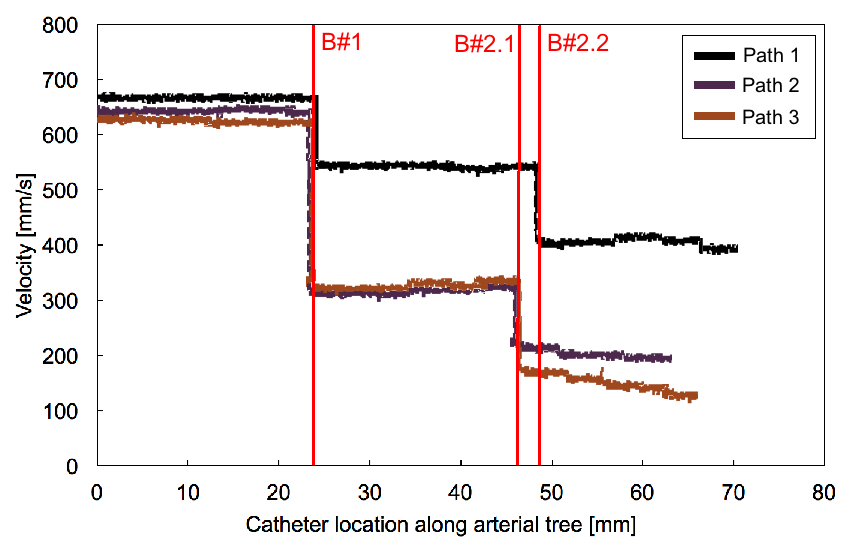}
  \vspace{0.2cm}
  \centerline{(b)}
\end{minipage}

\caption{(a) Arterial tree model with 4D flow streamlines of an anesthetized pig model with color-coded blood flow velocities. Symbols B$\#$ indicate bifurcations. (b) Mapping of blood flow velocities along various 3 vascular paths, based on tracked catheter location within the pig 's arterial tree model.}
\label{fig.Invivo}
\end{figure}

The final experiment consisted in an IRB-approved $in$ $vivo$ navigation with an anesthetized pig model. The pre-operative imaging was performed on a clinical 3T system (Achieva TX, Philips Healthcare, Best, The Netherlands), using a 16-channel thoracic surface coil for signal reception and the integrated 2-channel body coil for signal transmission. The field of view was of 240 x 300 x 80 mm, the acquired resolution 2.85 x 2.85 x 2.80 mm, the reconstructed resolution  1.35 x 1.35 x 1.4 mm, TR = 4.7 ms, TE = 2.7 ms, 8$^{\circ}$ flip angle, readout bandwidth of 498.4 Hz/pixel, SENSE acceleration factor of 2.5, a total of 25 reconstructed cardiac phases and velocity encoding (VENC) of 110 cm/s. Cardiac synchronization was performed using a peripheral pulse unit. Pre-injection acquisitions with respective flip angles of 4 and 20 degrees and the same acquisition parameters were also performed to enable the calculation of native T1 maps. For the clinical setup, only the OBR unit and laptop were required in the interventional suite. The experiment consisted in guiding the optical fiber triplet embedded in the catheter with 3 attempts from the femoral artery and into the arterial tree, each following distinct paths. Fig. \ref{fig.Invivo}a shows the representation of the arterial tree from the 4D-flow sequence. Fig. \ref{fig.Invivo}b presents the corresponding velocities  obtained from the 4D model along  each of the 3 paths of the the sensed catheter location during guidance. The results illustrate how the velocities drops once the catheter crosses bifurcation B$\#$1 into the common or splenic artery, as well as past B$\#$2.1 into the left or right branch or with  B$\#$2.2. This demonstrates the ability to locate the catheter in the arterial tree as it approaches vessel bifurcations.

\section{Conclusion}
We proposed a novel MR-compatible guidance system using an optical shape sensing catheter based on optical frequency domain reflectometry. Our system is the first to offer a fully distributed sensing device using Rayleigh scattering on UV exposed SMF fibers for navigation. In comparison to other single mode fibers, the UVE-SMF-28 allows to increase diffusion properties, leading to an improvement in tracking accuracy. Results show that this method offers tracking accuracies similar to theoretical estimations and EM tracking. Because the mapping is obtained with no user interaction using robust heat kernels to match curvature features, the proposed approach could be transposed to clinical practice for TACE of liver HCCs. Future work will improve the refresh rate with a high performance OBR (5Hz) and further experimentation with porcine models.
\\
\textbf{Acknowledgments:} We thank Drs. Guillaume Gilbert and An Tang for their contribution in the 4D-Flow sequence.

\bibliographystyle{splncs03}
\bibliography{sample}

\begin{thebibliography}{10}
\providecommand{\url}[1]{\texttt{#1}}
\providecommand{\urlprefix}{URL }

\bibitem{BADOUAL}
Badoual, A., Gerard, M., De~Leener, B., Abi-Jaoudeh, N., Kadoury, S.: {3D
  Vascular path planning of chemo-embolizations using segmented hepatic
  arteries from MR angiography}. IEEE ISBI pp. 225--228 (2016)

\bibitem{boscaini2016anisotropic}
Boscaini, D., Masci, J., Rodol{\`a}, E., et~al.: Anisotropic diffusion
  descriptors. In: Computer Graphics Forum. vol.~35, pp. 431--441. Wiley Online
  Library (2016)

\bibitem{duncan2007high}
Duncan, R.G., Froggatt, M.E., Kreger, S.T., et~al.: High-accuracy fiber-optic
  shape sensing. In: Int'l Symp. Smart Structures and Materials \&
  Nondestructive Evaluation and Health Monitoring. p. 65301S (2007)

\bibitem{Elayaperumal14}
Elayaperumal, S., Plata, J., Holbrook, A., et~al.: {Autonomous real-time
  interventional scan plane control with a 3-D shape-sensing needle}. IEEE
  Trans Med Imaging  33,  2128--39 (2014)

\bibitem{froggatt1998high}
Froggatt, M., Moore, J.: High-spatial-resolution distributed strain measurement
  in optical fiber with rayleigh scatter. Applied Optics  37(10),  1735--1740
  (1998)

\bibitem{froggatt2010fiber}
Froggatt, M.E., Duncan, R.G.: Fiber optic position and/or shape sensing based
  on rayleigh scatter (Aug~10 2010), {US Patent 7,772,541}

\bibitem{fuerst2016bioelectric}
Fuerst, B., Sutton, E.E., Ghotbi, R., et~al.: Bioelectric navigation: A new
  paradigm for intravascular device guidance. In: Proc. MICCAI. pp. 474--481
  (2016)

\bibitem{loranger2014rayleigh}
Loranger, S., Gagn{\'e}, M., Lambin-Iezzi, V., Kashyap, R.: Rayleigh scatter
  based order of magnitude increase in distributed temperature and strain
  sensing by simple uv exposure of optical fibre. Scientific reports  5,
  11177--11177 (2014)

\bibitem{mandal2016vessel}
Mandal, K., Parent, F., Martel, S., et~al.: Vessel-based registration of an
  optical shape sensing catheter for mr navigation. IJCARS  11(6),  1025--1034
  (2016)

\bibitem{Park10}
Park, Y.L., Elayaperumal, S., Daniel, B., et~al.: {Real-time estimation of 3-D
  needle shape and deflection for MRI-guided interventions}. IEEE/ASME Trans.
  Mechatronics  15(6),  906--915 (2010)

\bibitem{Roesthuis14}
Roesthuis, R., Kemp, M., van~den Dobbelsteen, J., Misra, S.: {Three-Dimensional
  Needle Shape Reconstruction Using an Array of Fiber Bragg Grating Sensors}.
  IEEE/ASME Trans. Mechatronics  19(4),  1115--1126 (2014)

\bibitem{song2014long}
Song, J., Li, W., Lu, P., et~al.: Long-range high spatial resolution
  distributed temperature and strain sensing based on optical frequency-domain
  reflectometry. IEEE Photonics Journal  6(3),  1--8 (2014)

\end{thebibliography}

\end{document}